\documentclass[aps,twocolumn,PRL,reprint,superscriptaddress]{revtex4}
\usepackage{amsmath}
\usepackage{graphicx}
\usepackage{bm}
\usepackage{amssymb}
\usepackage{hyperref}
\usepackage{xcolor}
\begin{document}
\title{Proposal for detecting nodal-line semimetal surface-states with resonant spin-flipped reflection}
\author{Wei Chen}
\email{pchenweis@gmail.com}
\affiliation{Institute for Theoretical Physics, ETH Zurich, 8093 Z\"{u}rich, Switzerland}
\affiliation{College of Science, Nanjing University of Aeronautics and Astronautics, Nanjing 210016, China}
\author{Kun Luo}
\affiliation{College of Science, Nanjing University of Aeronautics and Astronautics, Nanjing 210016, China}
\author{Lin Li}
\affiliation{College of Physics and Electronic Engineering, and Center for Computational Sciences, Sichuan Normal University, Chengdu, 610068, China}
\affiliation{Shenzhen Institute for Quantum Science and Engineering and Department of Physics, Southern University of Science and Technology, Shenzhen 518055, China}
\author{Oded Zilberberg}
\affiliation{Institute for Theoretical Physics, ETH Zurich, 8093 Z\"{u}rich, Switzerland}
\begin{abstract}
Topological nodal-line semimetals are predicted to exhibit unique drumhead-like surface states (DSS). Yet, a direct detection of such states remains a challenge. Here, we propose spin-resolved transport in a junction between a normal metal and a spin-orbit coupled nodal-line semimetal as the mechanism for their detection. Specifically, we find that in such an interface, the DSS induce resonant spin-flipped reflection. This effect can be probed by both vertical spin transport and lateral charge transport between anti-parallel magnetic terminals. In the tunneling limit of the junction, both spin and charge conductances exhibit a resonant peak around zero energy, providing a unique evidence of the DSS. This signature is robust to both dispersive-DSS and interface disorder. Based on numerical calculations, we show that the scheme can be implemented in the topological semimetal HgCr$_2$Se$_4$.
\end{abstract}

\maketitle

The discovery of topological materials has evinced one of the main recent advances in condensed matter physics~\cite{Hasan10rmp,Qi11rmp,Chiu16rmp}. Depending on whether the bulk states are gapped or gapless, topological materials can be largely divided into topological insulator phases \cite{Hasan10rmp,Qi11rmp} and topological semimetal phases \cite{Weng16jpcm}. In both categories, the material's bulk bands are characterized by topological invariants, which additionally result in gapless surface states according to a bulk-boundary correspondence \cite{Halperin82prb}. Therefore, detection of topological surface states is key for the identification of topological materials. For insulating phases, the edge/surface states are energetically well-separated from the bulk ones, and can be readily identified by transport measurement \cite{Konig07scn,Roth09scn}, scanning tunneling microscopy \cite{Roushan09nat,Zhang09prl} or angle-resolved photoemission spectroscopy (ARPES) \cite{Hsieh08nat}. Topological semimetals are more subtle, because the Fermi level crosses both the bulk and the surface states. Nevertheless, extensive progress has been achieved on the observation of exotic Fermi arc states in Weyl/Dirac semimetals \cite{Murakami07njp,Wan11prb,Weng15prx} by ARPES \cite{Liu14scn} and transport measurements \cite{Moll16nat}.

Recently, another kind of topological semimetal, nodal line semimetal (NLS), has attracted increasing research interests \cite{Burkov11prb,Kim15prl,Yu15prl,Heikkila11jetp,Weng15prb,Chen15nl,Zeng15arxiv,Fang15prb,Yamakage16jpsj,Xie15aplm,Chan16prb,Zhao16prb,Bian16prb,Bian16nc,Bzdusek16nat,Chenwei17prb,Yan17prb,Ezawa17prb}. These 3D materials are characterized by band crossings along closed loops, with each loop carrying a $\pi$ Berry flux~\cite{Burkov11prb}. A direct result of the NLS band-topology is the existence of drumhead-like surface states (DSS) nestled inside the projection of the nodal loops onto the 2D surface Brillouin zone \cite{Weng15prb}. There is a variety of candidates for NLS \cite{Kim15prl,Yu15prl,Heikkila11jetp,Weng15prb,Chen15nl,Zeng15arxiv,Fang15prb,Yamakage16jpsj,Xie15aplm,Chan16prb,Zhao16prb,Bian16prb,Bian16nc}, and their experimental characterization has seen recent progress using ARPES \cite{Bian16nc,Schoop16nc,Neupane16prb,Andreas16njp,Takane16prb} and quantum oscillation \cite{Hu16prl,Hu17prb,Kumar17prb,Pan17arxiv} measurements. However, a direct evidence of the novel DSS, the hallmark of NLS, is still missing: in the ARPES experiments, the surface states are veiled in the bulk bands, which can only be identified via a comparison with the results of a first-principles' calculation; the experiments on quantum oscillations only focus on bulk states, so that no information on the surface states can be extracted.

\begin{figure}
\centering
\includegraphics[width=0.48\textwidth]{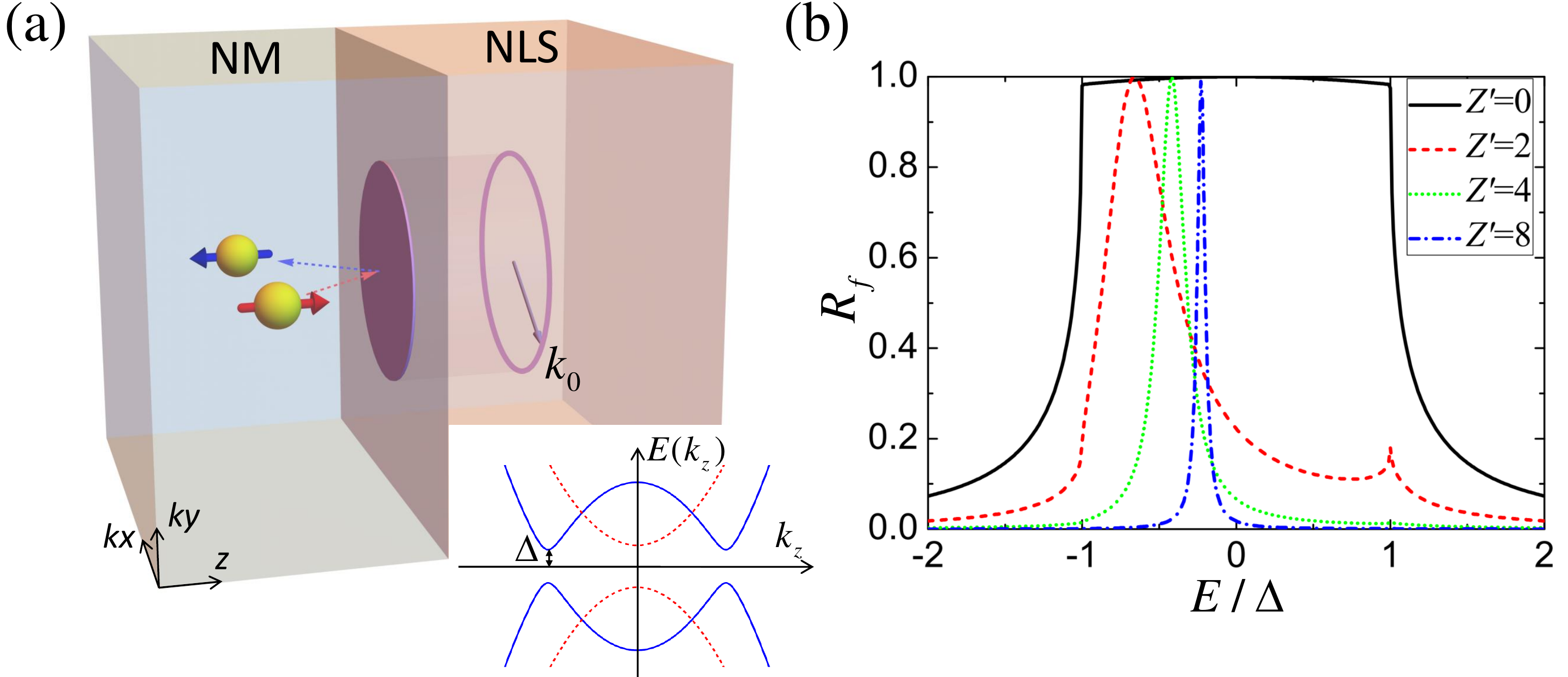}
\caption{(Color online). (a) Schematic illustration of a junction between a normal metal and a nodal line semimetal (NLS). The drumhead-like surface states at the boundary are encircled by the projection of the bulk nodal loop onto the surface Brillouin zone. Whenever, an incident electron from the metal is reflected, its spin is flipped when the transverse momentum lies inside the drumhead. The inset shows the $k_z$-dependent energy bands of effective 1D channels in the $z$-direction when the transverse momentum $\bm{k}_\parallel$ lies inside (blue solid lines) and outside (red dashed lines) the drumhead. (b) Probability of spin-flipped reflection at the boundary between the materials as a function of energy with following parameters: $\theta=\pi/6, k_F=1.1k_0=1.54, C=B=1, \lambda=0.01$, cf.~Eq.~\eqref{rf}. } \label{fig1}
\end{figure}

\begin{figure*}
\centering
\includegraphics[width=\textwidth]{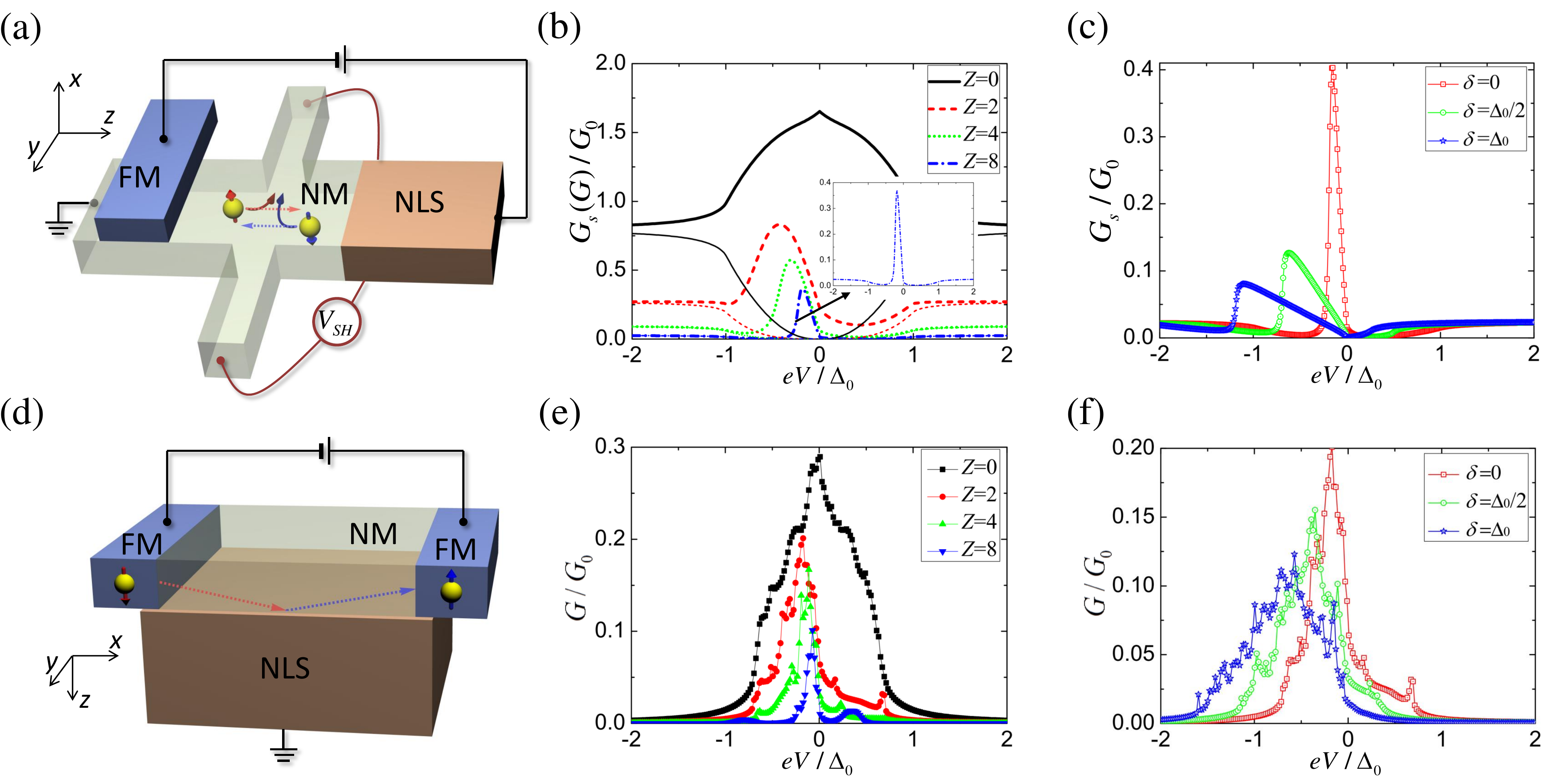}
\caption{(Color online). (a) Setup for spin transport DSS-detection. A normal metal with weak spin-orbit coupling, such as Al, Au, is fabricated into a Hall cross. Spin-up electrons are injected from a ferromagnetic metal (FM). The inverse spin Hall effect in the normal metal results in a transverse drift of electrons, yielding a transverse voltage drop $V_{SH}$ between two Hall probes~\cite{Valenzuela2006nat,Saitoh06apl}. (b) Spin (thick lines) and charge (thin lines) conductance for different interface barrier heights, cf.~Eq.~\eqref{Gs}. Inset: Zoom of the sharp peak with $Z=8$. (c) Effect of finite dispersion of the DSS on spin conductance with $Z=10$. (d) Setup for charge transport DSS-detection. The polarizations of the FM leads are taken to be antiparallel and transport through the device is possible only due to spin-flipping processes at the metal-NLS boundary. (e) Charge conductance for different interface barrier heights. (f) Effect of finite dispersion of the DSS on charge conductance with $Z=2$. All other parameters are the same as those taken in Fig. \ref{fig1}.} \label{fig2}
\end{figure*}

In this work, we propose two types of transport experiments for the detection of the DSS. These experiments rely on the spin-resolved scattering in a junction between a normal metal and a NLS, see Fig. \ref{fig1}(a). The DSS induce a resonant spin-flipped reflection (RSFR) for spin-polarized (along the $z$-axis) electrons incident from the normal metal. This effect manifests in a nearly-pure spin current flowing perpendicular to the junction [Fig. \ref{fig2}(a)], or in a lateral charge transport between two anti-parallel magnetic terminals [Fig. \ref{fig2}(d)]. In the tunneling limit, both setups show a resonant peak in their spin/charge conductances around the energy level of the nodal loop, that can serve as a direct evidence of the DSS. We analytically detail our predictions in a minimal NLS model and numerically demonstrate these signatures for a real material HgCr$_2$Se$_4$ \cite{Xu11prl}.

We consider spin-polarized electrons incident in the $z$-direction, see Fig. \ref{fig1}(a). We assume that the incident electron-spin is polarized in the $z$-direction and is injected into the metal ($z<0$) by a ferromagnetic lead. In the $z>0$ region, we use a minimal continuous model to describe the nodal-line semimetal as
\begin{equation}\label{HSM}
H_{\text{SM}}(\bm{k})=\lambda k_z\sigma_x+B(k_0^2-\left|\bm{k}\right|^2)\sigma_z,
\end{equation}
where $\left|\bm{k}\right|^2=k_x^2+k_y^2+k_z^2$ is the total momentum squared and the Pauli matrices $\sigma_{x,z}$ operate in the spin space. The Hamiltonian (\ref{HSM}) has eigenvalues $E_\pm=\pm\sqrt{\lambda^2k_z^2+B^2(k_0^2-\left|\bm{k}\right|^2)^2}$ and corresponding eigenstates $|u_\pm(\bm{k})\rangle$. The resulting two bands are degenerate at $k_{x}^2+k_{y}^2=k_0^2$ and $k_z=0$, thus defining a nodal loop in momentum space, see Fig. \ref{fig1}(a). Considering the transverse wavevector $\bm{k}_{\parallel}=(k_x,k_y)$ as a parameter, the Hamiltonian (\ref{HSM}) describes an effective 1D system in the $z$-direction. Whenever $\bm{k}_{\parallel}$ lies inside the nodal loop, that is $\left|\bm{k}_\parallel\right|<k_0$, the effective 1D system is insulating with an energy gap $\Delta(\bm{k}_\parallel)=\lambda k_0'$ opening around $k_z=k_0'=\sqrt{k_0^2-\left|\bm{k}_\parallel\right|^2}$, see inset of Fig. \ref{fig1}(a). The gap varies with $\bm{k}_\parallel$, and reaches its maximum $\Delta_0=\lambda k_0$ at $\bm{k}_\parallel=0$. Interestingly, the effective 1D model has a nontrivial band topology that is characterized by the Berry phase $\gamma_B=\pi$, with $\gamma_B=i\int_{-\infty}^\infty dk_z\langle u_-(\bm{k})|\partial_{k_z}|u_-(\bm{k})\rangle$ \cite{Chan16prb,Chiu16rmp}. In the presence of chiral symmetry, such nontrivial topological winding implies the appearance of a zero-energy end-state at an open boundary \cite{Ryu02prl,Hirayama17nc}. As $\bm{k}_\parallel$ varies inside the nodal loop, these topological end-states appear and form the DSS, which are encircled by the projection of the nodal loop onto the surface Brillouin zone, see Fig. \ref{fig1}(a). When the transverse wavevector $\bm{k}_\parallel$ lies outside the nodal loop, the effective 1D system becomes a trivial insulator with an energy gap around $k_z=0$ [inset of Fig. \ref{fig1}(a)], and no surface states show up at an open boundary.

In the $z<0$ region, lies the spin-degenerate normal metal, described by the Hamiltonian $H_{\text{NM}}=C\left|\bm{k}\right|^2-\mu_0$, where $C$ is a mass dependent parameter and $\mu_0$ is the chemical potential corresponding to the Fermi wavevector $\left|\bm{k}_F\right|=\sqrt{\mu_0/C}$. The interface scattering is considered using a Dirac-type barrier $U\delta(z)$. The scattering of the incident electron from the normal metal onto the NLS is solved by substituting $k_z=-i\partial_z$ and keeping $\bm{k}_{\parallel}$ a good quantum number~\cite{SM}. Importantly, we obtain that incident spin-up electrons with $|\bm{k}_\parallel|<k_0$ engender a spin-flipped reflection amplitude~\cite{SM}
\begin{equation}\label{rf}
r_f=-\frac{4}{(\eta+1/\eta+iZ')^2\Upsilon_1+(\eta-1/\eta+iZ')^2/\Upsilon_1},
\end{equation}
where $\eta=\sqrt{v_{\text{NM}}/v_{\text{SM}}}$ is the square root of the ratio of perpendicular velocities in the normal metal and NLS, with $v_{\text{NM}}=2C\left|\bm{k}_F\right|\cos\theta$ and $v_{\text{SM}}=2Bk_0'$, respectively, $\theta$ is the electron's incident angle (relative to the $z$-axis), $Z'=2U/\sqrt{v_{\text{NM}}v_{\text{SM}}}$ is a dimensionless interface barrier height, and $\Upsilon_{1}=[E+B(k_{1}^2-k_0'^2)]/(k_{1}\lambda)$ with $k_1=\sqrt{k_0'^2-(\lambda^2-\sqrt{4E^2B^2+\lambda^4-4\lambda^2B^2k_0'^2})/(2B^2)}$.

The obtained spin-flipped reflection probability, $R_f=|r_f|^2$, exhibits a sharp resonant peak around zero energy in the tunneling limit ($Z'\gg1$), see Fig. \ref{fig1}(b) \cite{note}. This result can be understood by rewriting Eq.~(\ref{rf}) in the tunneling limit as a summation over Feynman paths constructed by multiple reflection between the barrier and the NLS surface \cite{SM,Datta}. It turns out that the condition of RSFR is identical to the Bohr-Sommerfeld quantization condition for a surface bound state, which indicates that the RSFR is directly induced by the topological surface state.

Moreover, this scenario can also be understood by a tunneling Hamiltonian description. In the tunneling limit, for each 1D channel inside the nodal loop ($|\bm{k}_\parallel|<k_0$), we can use a tunneling Hamiltonian to describe the coupling between the DSS and the normal metal as
$H_{\text{T}}=\sum_{k_z}V_{\bm{k}}(c^\dag_{\bm{k}\uparrow}+ic^\dag_{\bm{k}\downarrow})\gamma_{\bm{k}_\parallel}+\text{H.c.}$,
where $\gamma_{\bm{k}_\parallel}=\int dz\frac{f^*_{\bm{k}_\parallel}(z)}{\sqrt{2}}\big[\psi_\uparrow(z)-i\psi_\downarrow(z)\big]$ is the Fermi operator for the DSS exhibiting some spatial distribution $f_{\bm{k}_\parallel}(z)$, $c_{\bm{k}\uparrow,\downarrow}$ are annihilation operators of electron in the normal metal, and $V_{\bm{k}}$ is the coupling strength~\cite{SM}. The DSS are spin-polarized along the $y$-direction~\cite{SM}, thus resulting in an equal coupling strength to both spin states in the normal metal. Such a tunneling Hamiltonian has the same form as that of a resonant tunneling through a single-level system~\cite{Ryndyk}, when we regard the two spin-states in the normal metal as two spinless leads and the surface state in each $\bm{k}_\parallel$-channel as the single-level. A direct calculation leads to a result of RSFR with a Lorentzian form, i.e., $R_f(E)=\Gamma^2/(E^2+\Gamma^2)$ ($\Gamma$ is level-width function)~\cite{SM}, which is consistent with the result in Fig. \ref{fig1}(b).

We propose two experimental schemes to probe the RSFR: (i) vertical spin transport in the setup in Fig. \ref{fig2}(a), and (ii) lateral charge transport in the setup in Fig. \ref{fig2}(d). For scheme (i), spin-polarized electrons are injected from a ferromagnetic lead, and then reflected with spin-flipping at the junction. The resulting nearly-pure spin current can be measured as a spin Hall voltage $V_{SH}$ \cite{Takahashi08jpsj} in the inverse spin Hall effect \cite{Valenzuela2006nat,Saitoh06apl}, see Fig. \ref{fig2}(a). For scheme (ii), charge current flows between two anti-parallel magnetic terminals, which cannot happen without the spin-flipped reflection. The RSFR can be well characterized in both setups by a resonant peak in the spin/charge conductances.

The spin current in setup (i) is defined as $I_s=I_\uparrow-I_\downarrow$, where $I_\sigma$ with $\sigma=\uparrow, \downarrow$ are spin-polarized currents flowing in the $z$-direction. In order to generate the spin Hall voltage, the spin is polarized along the $x$-direction, see Fig, \ref{fig2}(a). In the RSFR regime, incident and reflected electrons have opposite spin polarizations as well as opposite velocities. Consequently, the RSFR enhances the spin current, while the charge current $I=I_\uparrow+I_\downarrow$ is strongly suppressed. This results in a nearly-pure spin current flowing in the normal metal. To reveal the energy dependence of the spin transport, we calculate the differential spin conductance $G_s(eV)=\partial I_s/\partial V$ \cite{Mireles} using
\begin{equation}
G_s(E)=G_0\int_0^{\pi/2}d\theta\sin2\theta[1+R_f(\theta,E)-R_c(\theta,E)],
\label{Gs}
\end{equation}
where $G_0=\frac{\mathcal{A}k_F^2}{4\pi}\frac{e^2}{h}$ is the single-spin conductance of the uniform normal metal with a cross-section area $\mathcal{A}$, and $R_c=|r_c|^2$ is the probability of spin-conserved reflection.
The spin conductance $G_s$ as a function of the bias voltage $eV$ for different barrier strengths $Z=U/\sqrt{C|\bm{k}_F|Bk_0}$ is plotted in Fig. \ref{fig2}(b). In the transparent case ($Z=0$), $G_s$ exhibits a heightened ridge in the region $eV\in(-\Delta_0,\Delta_0)$, corresponding to strong spin-flipped reflection below the gap $\Delta_0$  [Fig. \ref{fig1}(b)]. As $Z$ increases, a narrower peak forms and moves towards zero energy, as expected for RSFR. At the same time, the height of the peak  reduces because the RSFR peaks become sharper in all transport channels [Fig. \ref{fig1}(b)], and the resonant energies do not match one another. Note that although the reduced spin conductance becomes small in the tunneling limit, the absolute value of $G_s$ around the resonant peak is still quite large. Concurrently, the charge conductance $G=\partial I/\partial V$ becomes much smaller than $G_s$ within the gap, indicating a high-purity spin current, see Fig. \ref{fig2}(b). Outside the gap, $G_s$ and $G$ tend to be equal, and transmission through the barrier (rather than RSFR) dominates the transport.

\begin{figure}
\centering
\includegraphics[width=0.48\textwidth]{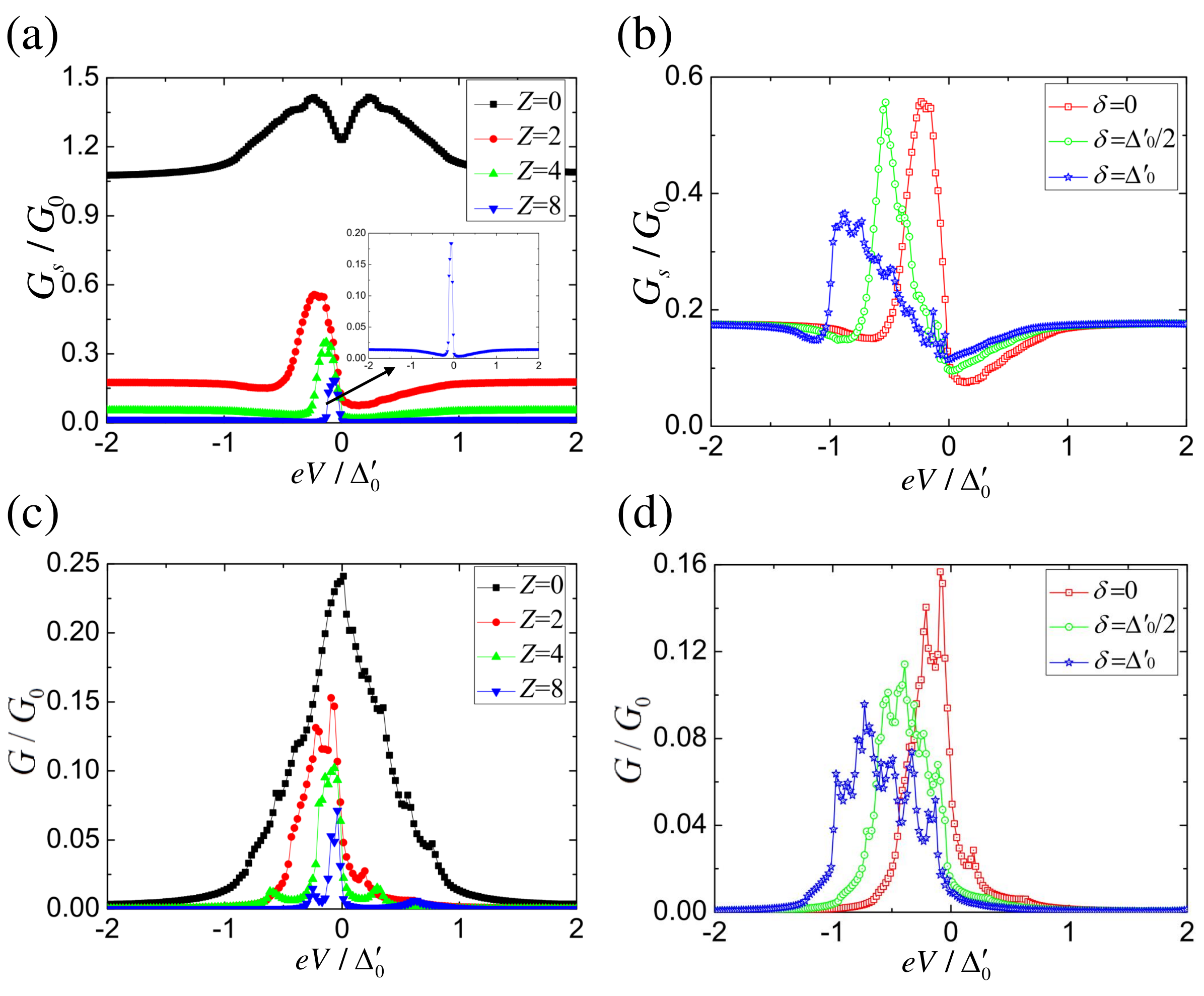}
\caption{(Color online). Numerical results for HgCr$_2$Se$_4$. Spin conductance of vertical spin transport for different (a) interface barriers, and (b) dispersions of DSS (with $Z=2$). Inset in (a): Zoom of the peak structure with $Z=8$. Charge conductance of lateral charge transport for different (c) interface barriers, and (d) dispersions of DSS (with $Z=2$). All the parameters are the same as those in Fig. \ref{fig1}, except that $D=0.01/k_0^2$.} \label{fig3}
\end{figure}

In real materials, chiral symmetry is usually broken on the open surface, and the DSS are commonly dispersive. To model this effect, we add a spin independent term $\varepsilon(\bm{k}_\parallel)=A _1(\left|\bm{k}_\parallel\right|^2-k_0^2)$ to Eq.~(\ref{HSM}). This introduces a band width $\delta=A_1k_0^2$ to the DSS. Such a $\bm{k}_\parallel$-dependent potential leads to further separation of RSFR levels in different channels. As a result, the peak of $G_s$ is broadened and also shifted, as shown in Fig. \ref{fig2}(c). In the tunneling limit, the width of the resonant peak is approximately equal to the width of the surface band $\delta$, so that the bandwidth of the DSS can be directly inferred from the width of the resonant peak in $G_s$ \cite{SM}, see also Figs.~\ref{fig2}(f) and \ref{fig3}(b,d).

The charge current in setup (ii) [Fig. \ref{fig2}(d)] flows in the normal metal in the $x$-direction, parallel to the interface of the junction. The normal metal is sandwiched by two anti-parallel ferromagnetic terminals. Without spin-flipped reflection at the junction, electrons injected from one terminal cannot enter the other. Therefore, setup (ii) can be used to detect the RSFR. The conductance $G$ is calculated numerically (using Kwant \cite{Groth14njp}) based on a lattice version of our model \cite{SM}, see Fig. \ref{fig2}(e). The conductance is normalized by $G_0$, the single-spin conductance in the $x$-direction through the normal metal. In the transparent limit of the junction ($Z=0$), electrons transport in the energy window $eV\in(-\Delta_0, \Delta_0)$, corresponding to the energy scale of spin-flipped reflection. As $Z$ increases, $G$ exhibits a sharp peak around zero energy, which signals the RSFR. In the setup in Fig. \ref{fig2}(d), multiple scattering occurs at all the surfaces of the normal metal, so that the conductance shows fluctuation. The effect of finite dispersion of the DSS is also investigated, and the results are shown in Fig. \ref{fig2}(f). It shifts and spreads the resonant peak, similar to the results reported for the spin transport in scheme (i), cf.~Fig. \ref{fig2}(c).

In realistic setups, there would be several additional imperfections that should be taken into account, such as interface imperfections and nonpure spin injection \cite{Stroud02prl}. In experiments, interface imperfections commonly exist, such that the clean tunneling limit is difficult to obtain. For both transport schemes, we numerically investigate this effect by introducing interface disorder (see Fig. S.2 in the Supplemental Material~\cite{SM}). One can see that apart from some broadening of the general features, the resonant peak in the spin/charge conductances are robust to strong disorder with the strength close to the interface barrier, reflecting the robustness of topological DSS. Similarly, spin-polarization averaging leads to an overall reduction prefactor that does not qualitatively change the overall transport signatures~\cite{SM}.

Our analysis has, thus far, relied on a minimal NLS model (\ref{HSM}). For experimental realizations, we consider the topological semimetal HgCr$_2$Se$_4$ \cite{Xu11prl} as a promising candidate. In the $|J,M_J\rangle$ basis $|\frac{3}{2}, \frac{3}{2}\rangle=\frac{1}{\sqrt{2}}|(X+iY)\uparrow\rangle$ and $|S, \downarrow\rangle$, an effective two-band model for HgCr$_2$Se$_4$ can be written as
\begin{equation}\label{HHCS}
\begin{split}
H_{\text{HCS}}(\bm{k})&=\left(
          \begin{array}{cc}
            B(k_0^2-\left|\bm{k}\right|^2) & Dk_zk_-^2 \\
            Dk_z k_+^2 & -B(k_0^2-\left|\bm{k}\right|^2) \\
          \end{array}
        \right),
\end{split}
\end{equation}
with $k_\pm=k_x\pm ik_y$. The eigenenergies of this model are $E'_\pm=\pm\sqrt{B^2(k_0^2-\left|\bm{k}\right|^2)^2+D^2k_z^2\left|\bm{k}_\parallel\right|^4}$. Therefore, the gap closes along exactly the same nodal line as that of the minimal model (\ref{HSM}). For each $\bm{k}_\parallel$-channel, the effective gap is $\Delta'(\bm{k}_\parallel)=Dk_0'\left|\bm{k}_\parallel\right|^2$, and its maximum is $\Delta'_0=2Dk_0^3/(3\sqrt{3})$ when $|\bm{k}_\parallel|=\sqrt{2/3}k_0$. Additionally, the model \eqref{HHCS} exhibits two Weyl nodes in the $z$-axis at $k_z=\pm k_0$. The Weyl nodes only introduce a single gapless 1D channel, and the corresponding Fermi-arc surface states do not appear at an open boundary in the $z$-direction, so that the DSS remains the dominating transport effect at a metal-NLS junction in Fig. \ref{fig1}(a).

We numerically calculate the spin/charge conductances in schemes (i) and (ii) [Figs. \ref{fig2}(a) and \ref{fig2}(d)] for lattice version of Eq.~\eqref{HHCS}~\cite{SM}. All the results agree well with those of the minimal model. $G_s$ and $G$ for different barrier heights are shown in Figs. \ref{fig3}(a) and \ref{fig3}(c). For a transparent junction, $G_s$ is heightened and $G$ has a peak spreading in the energy window $eV\in(-\Delta'_0, \Delta'_0)$, which is generated by the spin-flipped reflection. As $Z$ increases, a resonant peak shows up, indicating the DSS induced RSFR. The effect of dispersive-DSS is shown in Figs. \ref{fig3}(b) and \ref{fig3}(d) and leads to widening of the RSFR peak. The resonant peak is robust against interface disorder and nonpure spin injection~\cite{SM}.

It is worthwhile to compare the DSS induced RSFR with other spin relaxation processes. Most spin relaxation mechanisms can only lead to weak dissipation of spin signatures \cite{Sarma04rmp}, strongly different from the RSFR-induced enhancement of the  spin signature. Consider, for example, spin-flipped scattering induced by magnetic impurities at the interface of the junction: in order to obtain a comparable resonance strength, a high density of impurities with the same energy level is required. Similarly, the electrons will have a very small rate of colliding with bulk impurities due to the vanishing density of states in the bulk of the NLS.

%In contrast with the DSS, common surface states in noble metals cannot induce RSFR, as the surface states have the following properties: (i) very strong dispersion, as well as a huge density of states from the bulk, excluding a resonant transport peak; (ii) two branches with opposite spin polarization in Rashba-type surface states tend to cancel out the spin-flipping effect. Furthermore, the common surface states are sensitive to the surface disorder, and can be easily pushed to the bulk continuum spectrum.

To conclude, we have shown that resonant spin-flipped reflection can serve as an unambiguous evidence of the drumhead-like surface states in the spin-orbit coupled nodal-line semimetal. Recent experimental progress on spin-resolved transport in HgCr$_2$Se$_4$ \cite{Guan15prl} paves the way to the realization of our proposal. Our analysis can be extended to other types of nodal-line semimetals, i.e., both to additional materials but also engineered systems such as photonic nodal-line systems.

\begin{acknowledgments}
We thank Chen Fang, Gang Xu, Wei-Yin Deng, Di Wu, Zhong Wang, J. L. Lado, and Michael S. Ferguson for helpful discussions and Hu Zhao for assistance on the figures. This work was supported by the National Natural Science Foundation of China under Grants No. 11504171 (W.~C.), No. 11604138 (L.~L.) and by the Natural Science Foundation of Jiangsu Province in China under Grant No. BK20150734. W.~C. acknowledges the support from the Swiss Government Excellence Scholarship under the program of China Scholarships Council (No. 201600160112). O.~Z. acknowledges financial support from the Swiss National Science Foundation.
\end{acknowledgments}

%\bibliographystyle{apsrev4-1-etal-title}
%\bibliography{refs-toposemimetal}

\begin{thebibliography}{48}
\bibitem{note} The resonant peak slightly deviates from zero energy, because the Dirac-type potential does not
have  particle-hole  symmetry,  i.e.,  it  has  the  same  sign  for  both  spin-up  and  spin-down  states. Whether the peak is on the left or right side of zero energy depends on the sign of the barrier strength $Z$. Here, we have taken the bands in the normal metal with a positive effective mass, so that $Z$ is positive. If we would have used the bands with negative mass, then the barrier $Z$ for the hole-like particle is also negative, and the corresponding peak will appear on the right side of the zero energy.
\end{thebibliography}


\newpage
\onecolumngrid
\renewcommand{\theequation}{S.\arabic{equation}}
\setcounter{equation}{0}
\renewcommand{\thefigure}{S.\arabic{figure}}
\setcounter{figure}{0}

\section{Supplemental Material for ``Proposal for detecting nodal-line semimetal surface-states with resonant spin-flipped reflection''}

\subsection{Effective continuous  model and scattering approach}
In order to solve the scattering problem in a junction between a normal metal and a NLS, we write the Hamiltonian of the whole system in real space in the $z$-direction. Substituting $k_z=-i\partial_z$ and keeping $\bm{k}_{\parallel}$ as a parameter, the system can be described by an effective 1D Hamiltonian
\begin{equation}\label{HSM2}
\begin{split}
\mathcal{H}(z,\bm{k}_{\parallel})&=-\partial_z C(z)\partial_z-\mu(z,\bm{k}_{\parallel})+U\delta(z)\\
&-\frac{i}{2}\{\lambda(z),\partial_z\}\sigma_x+[B(z)k_0'^2+\partial_zB(z)\partial_z]\sigma_z
\end{split}
\end{equation}
where $C(z)=C\theta(-z)$, $\mu(z,\bm{k}_{\parallel})=(\mu_0-C|\bm{k}_{\parallel}|^2)\theta(-z)$ is the effective chemical potential, $\lambda(z)=\lambda\theta(z)$, $B(z)=B\theta(z)$, and $U\delta(z)$ is the interface barrier of Dirac type. Note that all the parameters are spatially varying. In order to keep the Hermiticity of the Hamiltonian, all terms in Eq. (\ref{HSM2}) are symmetrized (cf.~Refs.~\cite{BenDaniel66pr,Zulicke02prl} for more details on the method). A solution, $\psi$, to this scattering problem obeys certain boundary conditions in the $z$-direction. Assuming $\psi$ to be continuous at the interface $z=0$, and integrating Eq. (\ref{HSM2}) across the interface, one reaches the following boundary conditions
\begin{equation}\label{boundary}
\begin{split}
\psi(0^+)&=\psi(0^-), \\
B\sigma_z\psi'(0^+)+C\psi'(0^-)&=(\frac{i}{2}\lambda\sigma_x-U)\psi(0).
\end{split}
\end{equation}
Such boundary conditions guarantee current conservation in the $z$-direction at the interface, that is, $\mathcal{J}_{\text{SM}}=\mathcal{J}_{\text{NM}}$, where the currents in the normal metal and NLS are $\mathcal{J}_{\text{NM}}=2C\text{Im}(\psi^\dag\partial_z\psi)$ and $\mathcal{J}_{\text{SM}}=\lambda\psi^\dag\sigma_x\psi-2B\text{Im}(\psi^\dag\sigma_z\partial_z\psi)$, respectively.

For a spin-up (in the $z$-direction) electron with $|\bm{k}_\parallel|<k_0$ incident from the normal metal ($z<0$) with an incident angle $\theta$ (relative to the $z$-axis), the scattering states in the two regions are
\begin{equation}\label{wave}
\begin{split}
\psi(z\leq0)&=\Big[\left(
                   \begin{array}{c}
                     1 \\
                     0 \\
                   \end{array}
                 \right)e^{ik_\perp  z}
+\left(
             \begin{array}{c}
               r_c \\
               r_f \\
             \end{array}
           \right)e^{-ik_\perp  z}\Big]e^{ik_\parallel r_\parallel}\\
\psi(z>0)&=\Big[t_1\left(
             \begin{array}{c}
               u_1 \\
               v_1 \\
             \end{array}
           \right)e^{ik_1 z}
+t_2\left(
             \begin{array}{c}
               u_2 \\
               v_2 \\
             \end{array}
           \right)e^{ik_2 z}\Big]e^{ik_\parallel r_\parallel},
\end{split}
\end{equation}
where the perpendicular and parallel wave vectors of the incident electron are $k_\perp=|\bm{k}_F| \cos\theta$ and $k_\parallel=|\bm{k}_F|\sin\theta$ with $\bm{k}_F$ the Fermi wavevector in the normal metal and $|\bm{k}_F|=\sqrt{\mu_0/C}$. The transmitted wavefunctions are defined through $\Upsilon_{1,2}=v_{1,2}/u_{1,2}=[E+B(k_{1,2}^2-k_0'^2)]/(k_{1,2}\lambda)$, and the wavenumbers $k_{1,2}$ in the $z$-direction are determined by $k_{1,2}^2=k_0'^2-(\lambda^2\mp\sqrt{4E^2B^2+\lambda^4-4\lambda^2B^2k_0'^2})/(2B^2)$. The signs of the wave vectors are chosen such that the transmission waves either propagate freely or decay in the $z$-direction, depending on whether the incident energy is above or below the energy gap. Scattering amplitudes $r_c$ and $r_f$ refer to spin-conserved and spin-flipped reflection, respectively, and $t_{1,2}$ are the transmission amplitudes.

Generally, by inserting Eq.~\eqref{wave} into Eq.~\eqref{boundary}, we obtain an analytical solution for all the scattering amplitudes. The expressions, however, are quite unwieldy and to obtain a simple solution, we first take the limit $\lambda\ll Bk_0'$. Then $\Upsilon_1\Upsilon_2=-1$, and the second boundary condition in Eq. (\ref{boundary}) reduces to $B\sigma_z\psi'(0^+)+C\psi'(0^-)=U\psi(0)$. Inserting the wave function (\ref{wave}) into the boundary conditions, we obtain the amplitude of spin-flipped reflection
\begin{equation}
r_f=-\frac{4}{(\eta+1/\eta+iZ')^2\Upsilon_1+(\eta-1/\eta+iZ')^2/\Upsilon_1},
\end{equation}
where $\eta=\sqrt{v_{\text{NM}}/v_{\text{SM}}}$ is the square root of the ratio of perpendicular velocities in the normal metal and NLS, with $v_{\text{NM}}=2C|\bm{k}_F|\cos\theta$ and $v_{\text{SM}}=2Bk_0'$, respectively, and $Z'=2U/\sqrt{v_{\text{NM}}v_{\text{SM}}}$ is the dimensionless barrier strength. The amplitude $r_c$ of spin-conserved reflection can be similarly obtained.

\subsection{Feynman path explanation}

\begin{figure}
\centering
\includegraphics[width=0.6\textwidth]{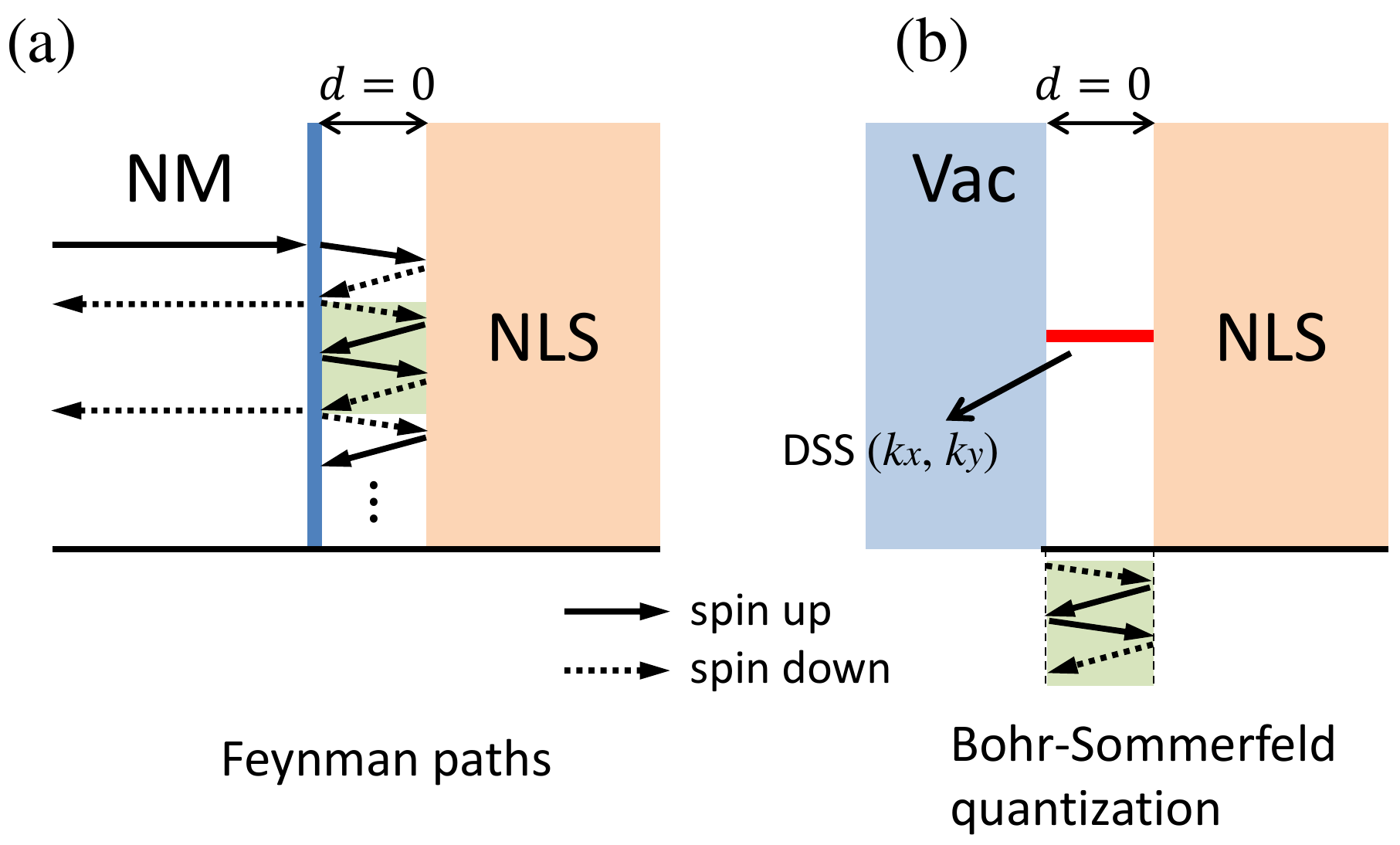}
\caption{(Color online). Correspondence between DSS and RSFR.  The solid and dashed lines with arrows denote the propagations of spin-up and spin-down electrons, respectively. (a) RSFR interpreted by Feynman path summation. Trajectories in the dashed area form the basic scattering loop. (b) Topological end state [DSS ($k_x, k_y$)] can be understood as a bound state (red line segment) confined by the NLS and the vacuum (Vac). Trajectories in the dashed area satisfy the Bohr-Sommerfeld quantization condition for the bound state.} \label{figs1}
\end{figure}

DSS induced RSFR can be understood by rewriting $r_f$ as a summation over Feynman paths. In the tunneling limit ($Z'\gg1$), the value of $\eta$ is not important, so that it can be set to unity for simplicity. The amplitude of spin-flipped reflection reduces to
\begin{equation}\label{rf1}
r_f=-\frac{4}{(2+iZ')^2\Upsilon_1-Z'^2/\Upsilon_1}.
\end{equation}
This result can also be obtained by a Feynman path summation over all possible scattering processes between the barrier and the surface of the NLS, as shown in Fig. \ref{figs1}(a). The barrier lies inside the normal metal and is infinitesimally close to the NLS. The lowest order contribution to $r_f$ is composed of two transmission processes at the barrier and one spin-flipped reflection at the NLS surface. Higher-order contributions contain an integer number of basic scattering loops, as sketched in Fig. \ref{figs1}(a). The final amplitude of spin-flipped reflection is a summation over all of these processes, yielding
\begin{equation}\label{rf2}
r_f=t^2r_f^0\big[1+r^2r^0_f\tilde{r}^0_f+(r^2r^0_f\tilde{r}^0_f)^2+\cdots\big]=\frac{t^2r_f^0}{1-r^2r^0_f\tilde{r}^0_f},
\end{equation}
where the amplitudes of spin-flipped reflection $r_{f}^0=-1/\Upsilon_1=-e^{-i\cos^{-1}(E/\Delta)}$ and $\tilde{r}_{f}^0=-r_{f}^0$ correspond to spin-up and spin-down incident electrons in the transparent barrier limit ($Z'=0, \eta=1$); $t=\sqrt{T}e^{i\alpha}=[1+iZ'/2]^{-1}$ and $r=-\sqrt{R}e^{i\beta}=t-1$ are amplitudes of transmission and reflection by the barrier in the normal metal. Inserting all of the parameters into Eq. (\ref{rf2}), we obtain the same result as Eq. (\ref{rf1}).

The advantage of using the expression (\ref{rf2}) is that the energy level of the RSFR can be easily extracted, by setting the denominator of $r_f$ to zero, i.e., by imposing $r^2r^0_f\tilde{r}^0_f=1$. For $Z'\gg1$, the condition for RSFR reduces to $\cos^{-1}(E/\Delta)=\beta+\pi/2$. As the barrier height increases to infinity, the phase $\beta$ decreases to zero, leading to a zero-energy resonance.

Note that the condition of RSFR is identical to the Bohr-Sommerfeld quantization condition for a surface bound state, $\text{Arg}(r^2r_f^0\tilde{r}_f^0)=2n\pi$, see Fig. \ref{figs1}(b). Such correspondence between the RSFR and the periodic boundary condition for the bound state indicates that the RSFR is directly induced by the topological surface states, thus providing an effective way to detect the DSS.

\subsection{Tunneling Hamiltonian description of RSFR induced by DSS}
In this section, we present a tunneling Hamiltonian description of DSS induced RSFR. This method depends only on the low energy physics of the DSS, irrespective of the specific physical system, i.e., it reveals the universality of the DSS induced RSFR.

First, we solve the zero-energy end state localized around the open boundary $z=0$ for the semi-infinite NLS ($z>0$). Taking $k_x, k_y$ as parameters, the Hamiltonian in Eq. (1) in the main text can be interpreted as an effective 1D Hamiltonian in the $z$-direction as
\begin{equation}
\mathcal{H}_{\text{SM}}(z,\bm{k}_\parallel)=\lambda k_z\sigma_x+B(k_0'^2-k_z^2)\sigma_z.
\end{equation}
Substituting $k_z=-i\partial_z$, we obtain the equation for the zero-energy wave function
\begin{equation}
\big[-i\lambda\partial_z\sigma_x+B(k_0'^2+\partial_z^2)\sigma_z\big]\phi_{\bm{k}_\parallel}(z)=0.
\end{equation}
Multiplying $\sigma_x$ from the left-hand side, we obtain
\begin{equation}
\partial_z\phi_{\bm{k}_\parallel}(z)=-\frac{B}{\lambda}(k_0'^2+\partial_z^2)\sigma_y\phi_{\bm{k}_\parallel}(z).
\end{equation}
We choose the wave function to be an eigenstate of $\sigma_y$, that is $\phi_{\bm{k}_\parallel}(z)=\chi_\eta f_{\bm{k}_\parallel}(z)$, with $\sigma_y\chi_\eta=\eta\chi_\eta(\eta=\pm1)$. Then the differential equation reduces to
\begin{equation}
\partial_zf_{\bm{k}_\parallel}(z)+\frac{\eta B}{\lambda}(k_0'^2+\partial^2_z)f_{\bm{k}_\parallel}(z)=0.
\end{equation}
Taking the trial wavefunction $f_{\bm{k}_\parallel}(z)\propto e^{-\kappa z}$,  yields the secular equation
\begin{equation}
B\kappa^2-\eta\lambda\kappa+Bk_0'^2=0.
\end{equation}
The two roots of the secular equation satisfy $\kappa_++\kappa_-=\eta\lambda/B$ and $\kappa_+\kappa_-=k_0'^2$. The boundary conditions for the end state are
\begin{equation}
f_{\bm{k}_\parallel}(z=0)=0,\ \ f_{\bm{k}_\parallel}(z=+\infty)=0.
\end{equation}
The second boundary condition requires that both roots should be positive, that is $\kappa_\pm>0$, so that the wavefunction decays to zero at infinity. Taking $\lambda, B$ to be positive, that means $\eta=+1$, and $k_0'^2=k_0^2-|\bm{k}_\parallel|^2>0$. Therefore, the end state has a spin polarization along the $y$-direction. This procedure can be applied to all $\bm{k}_\parallel$ channels, i.e., to all states that are encircled by the projection of the nodal loop onto the surface Brillouin zone, leading to the conclusion that the DSS are spin-polarized in the $y$-direction. With the help of the first boundary condition, the wave function of the zero-energy end state can be obtained as
\begin{equation}
\phi_{\bm{k}_\parallel}(z)=\frac{f_{\bm{k}_\parallel}(z)}{\sqrt{2}}\left(
                              \begin{array}{c}
                                1 \\
                                i \\
                              \end{array}
                            \right),
\end{equation}
with $f_{\bm{k}_\parallel}(z)=C_0(e^{-\kappa_+z}-e^{-\kappa_-z})$ being the normalized spatial wavefunction, and $\kappa_\pm=\lambda/2B\pm\sqrt{(\lambda/2B)^2-k_0'^2}$.

Now we are ready to introduce the tunneling Hamiltonian describing the coupling between the normal metal and the DSS. As long as $|\bm{k}|^2<k_0^2$, there exists a zero-energy end state for this channel. The corresponding Fermi operator can be defined as
\begin{equation}
\gamma^\dag_{\bm{k}_\parallel}=\int dz\frac{f_{\bm{k}_\parallel}(z)}{\sqrt{2}}\big[\psi^\dag_\uparrow(z)+i\psi^\dag_\downarrow(z)\big],
\end{equation}
and the Hamiltonian for the end state ($E_0=0$) is
\begin{equation}
H_0=E_0\gamma_{\bm{k}_\parallel}^\dag\gamma_{\bm{k}_\parallel}.
\end{equation}
The Hamiltonian for the transport channel ($k_x, k_y$) in the normal metal is
\begin{equation}
H_{\text{NM}}=\sum_{k_z,\sigma=\uparrow\downarrow}\varepsilon_{\bm{k}}c^\dag_{\bm{k}\sigma}c_{\bm{k}\sigma},
\end{equation}
with $\varepsilon_{\bm{k}}=C|\bm{k}|^2-\mu_0$.
The coupling between the end state and the normal metal can be described by
\begin{equation}
H_{\text{T}}=\sum_{k_z,\sigma=\uparrow\downarrow}\int dz\big[t_{\bm{k}}(z)c^\dag_{\bm{k}\sigma}\psi_\sigma(z)+\text{H.c.}\big]
\end{equation}
Since we focus on low-energy scales well below the gap $\Delta(\bm{k}_\parallel)=\lambda k_0'$, we can project the field operator $\psi(z)=(\psi_\uparrow,\psi_\downarrow)^{\text{T}}$ onto the zero-energy bound state as
\begin{equation}
\psi(z)\approx\gamma_{\bm{k}_\parallel}\phi_{\bm{k}_\parallel}(z).
\end{equation}
Then the tunneling Hamiltonian reduces to
\begin{equation}
H_{\text{T}}=\sum_{k_z}V_{\bm{k}}(c^\dag_{\bm{k}\uparrow}+ic^\dag_{\bm{k}\downarrow})\gamma_{\bm{k}_\parallel}+\text{H.c.},
\end{equation}
where $V_{\bm{k}}=\int dz t_{\bm{k}}(z)f_{\bm{k}_\parallel}(z)/\sqrt{2}$.

The full retarded and advanced Green's function for the end state can be solved based on the tunneling model, and the result is
\begin{equation}
G_\gamma^{R/A}=\frac{1}{\omega-E_0-\Sigma^{R/A}_\uparrow-\Sigma^{R/A}_\downarrow},
\end{equation}
with the self-energies contributed by both spin states in the normal metal being
\begin{equation}
\Sigma^{R/A}_\uparrow=\Sigma^{R/A}_\downarrow=\sum_{k_z}\frac{|V_{\bm{k}}|^2}{\omega-\varepsilon_{\bm{k}}\pm i\delta}.
\end{equation}
Given that the coupling strength $V_{\bm{k}}$ is slowly varying with $k_z$, we can neglect the real part of $\Sigma^{R/A}_{\uparrow,\downarrow}$, so that
\begin{equation}
\Sigma^{R/A}_{\uparrow,\downarrow}\simeq\mp i\sum_{k_z}\pi|V_{\bm{k}}|^2\delta(\omega-\varepsilon_{\bm{k}})=\mp i\Gamma_{\uparrow,\downarrow}/2,
\end{equation}
where $\Gamma_{\uparrow}=\Gamma_{\downarrow}=\Gamma$ is the  level-width function.
The probability of spin-flipped reflection can be obtained by
\begin{equation}
R_f(E)=\Gamma_\uparrow G^A_\gamma(E) \Gamma_\downarrow G^R_\gamma(E)=\frac{\Gamma^2}{(E-E_0)^2+\Gamma^2}.
\end{equation}
Because the coupling strengths between the end state and both spin states are equal, the spin-flipped reflection probability has a Lorentzian form, with a resonance at $E=E_0=0$. If we keep the small real part of the self-energy, then the resonance will deviate from zero energy by a small value. In the tunneling limit $\Gamma\rightarrow0$, the resonant energy level tends to zero energy, which is consistent with Fig. 1(b) in the main text. Such a resonance occurs for all $(k_x, k_y)$ channels, contributing to the final resonant peak in the conductance spectra.

Based on the tunneling Hamiltonian description, one can see that the result of DSS induced RSFR is quite general, independent on the  type of the nodal line and the specific distribution of the DSS in the surface Brillouin zone.

\subsection{Effect of surface dispersion}
In real materials of NLS, the DSS are commonly dispersive. To include this effect, we add a spin-independent term $\varepsilon(\bm{k}_\parallel)$ to the Hamiltonian of the NLS. Such a term is nothing but a $\bm{k}_\parallel$-dependent potential. As a result, the energy of the end state in each $\bm{k}_\parallel$-channel is shifted from zero by $\varepsilon(\bm{k}_\parallel)$. Here, we investigate the effect of finite dispersion of the DSS on the vertical spin transport scheme [Fig. 2(a) in the main text]. In the tunneling limit, the spin-relevant transport is dominated by DSS induced RSFR below the gap $\Delta_0$. The spin conductance can be evaluated by
\begin{equation}
G_s(E)=\frac{e^2}{h}\frac{2\mathcal{A}}{(2\pi)^2}\int\int_{|\bm{k}_\parallel|<k_0} dk_xdk_yR_f\big(E-\varepsilon(\bm{k}_\parallel)\big).
\end{equation}
Rewriting the integral over energy, yields
\begin{equation}
G_s(E)=\frac{2e^2}{h}\int d\varepsilon N(\varepsilon) R_f\big(E-\varepsilon\big),
\end{equation}
where $N(\varepsilon)=\frac{\mathcal{A}}{2\pi} \frac{|\bm{k}_\parallel|(\varepsilon)}{\partial\varepsilon/\partial |\bm{k}_\parallel|}$ is the density of DSS. We assume that the DSS have finite density of states within the energy interval $\varepsilon\in[\varepsilon_1, \varepsilon_2]$. In the tunneling limit, $R_f(E)$ possesses a narrow peak structure around zero energy, cf. Fig. 1(b) in the main text. Therefore, the above expression has a considerable weight only in the energy window $E\in[\varepsilon_1, \varepsilon_2]$, which means the width of the resonant peak is approximately equal to the width of the surface band, $\delta=\varepsilon_2-\varepsilon_1$. In the main text, we adopt a specific example $\varepsilon(\bm{k}_\parallel)=A _1(\left|\bm{k}_\parallel\right|^2-k_0^2)$ to simulate this effect, with the corresponding density of DSS, $N(\varepsilon)=\mathcal{A}/(4\pi A_1)$, and $\varepsilon_1=-\delta, \varepsilon_2=0$. Consequently, the spin conductance reads as
\begin{equation}
G_s=\frac{e^2}{h}\frac{\mathcal{A}}{2\pi A_1}\int^0_{-\delta}d\varepsilon R_f(E-\varepsilon).
\end{equation}
The resulting $G_s$ has considerable contributions only in the energy interval $E\in[-\delta, 0]$, which characterizes the width of the resonant peak, consistent with the numerical results in Figs. (2) and (3) in the main text.

\subsection{Lattice model and numerical simulation}
The numerical calculations reported in the main text are performed using KWANT\cite{Groth14njp} and based on lattice versions of the main models in the main text. For the minimal model in Eq.~(1) in the main text, the mapping to a cubic lattice is obtained by substituting $k_i\rightarrow\frac{1}{a}\sin k_ia$ and $k_i^2\rightarrow\frac{2}{a^2}(1-\cos k_ia)$. Performing Fourier transformation, we obtain the following tight-binding Hamiltonian
\begin{equation}
\begin{split}
H_{\text{SM}}^{\text{latt}}
&=\sum_{i}B(k_0^2-\frac{6}{a^2})c_i^\dag\sigma_zc_i+A_1(\frac{4}{a^2}-k_0^2)c_i^\dag c_i\\
&+\sum_{i}\frac{\lambda}{2a}e^{-\frac{\pi}{2}i}c_i^\dag\sigma_xc_{i+a\hat{z}}+\text{H.c.}\\
&+\sum_{i}\frac{B}{a^2}(c_i^\dag\sigma_zc_{i+a\hat{x}}+c_i^\dag\sigma_zc_{i+a\hat{y}}+c_i^\dag\sigma_zc_{i+a\hat{z}})+\text{H.c.}\\
&-\sum_{i}\frac{A_1}{a^2}(c_i^\dag c_{i+a\hat{x}}+c_i^\dag c_{i+a\hat{y}})+\text{H.c.}.
\end{split}
\end{equation}

Similarly, the lattice model for the normal metal is
\begin{equation}
\begin{split}
H_{\text{NM}}^{\text{latt}}
&=\sum_i(\frac{6C}{a^2}-\mu_0)c_i^\dag c_i\\
&-\sum_i\frac{C}{a^2}(c^\dag_ic_{i+a\hat{x}}+c^\dag_ic_{i+a\hat{y}}+c^\dag_ic_{i+a\hat{z}})+\text{H.c.},
\end{split}
\end{equation}
and the lattice Hamiltonian for the HgCr$_2$Se$_4$ model in Eq. (4) in the main text becomes
\begin{equation}
\begin{split}
H_{\text{HCS}}^{\text{latt}}&=\sum_i \frac{D}{2a^3}[e^{-(\pi/2)i}c^\dag_i\sigma_x c_{i+a\hat{y}+a\hat{z}}+e^{-(\pi/2)i}c^\dag_i\sigma_x c_{i-a\hat{y}+a\hat{z}}\\
&+e^{(\pi/2)i}c^\dag_i\sigma_x c_{i+a\hat{x}+a\hat{z}}+e^{(\pi/2)i}c^\dag_i\sigma_x c_{i-a\hat{x}+a\hat{z}}]+\text{H.c.}\\
&+\sum_i\frac{D}{4a^3}[e^{(\pi/2)i}c_i^\dag\sigma_yc_{i+a\hat{x}+a\hat{y}+a\hat{z}}+e^{(\pi/2)i}c_i^\dag\sigma_yc_{i-a\hat{x}-a\hat{y}+a\hat{z}}\\
&+e^{(\pi/2)i}c_i^\dag\sigma_yc_{i-a\hat{x}+a\hat{y}-a\hat{z}}+e^{(\pi/2)i}c_i^\dag\sigma_yc_{i+a\hat{x}-a\hat{y}-a\hat{z}}]+\text{H.c.}\\
&+\sum_i\frac{B}{a^2}(c_i^\dag\sigma_zc_{i+a\hat{x}}+c_i^\dag\sigma_zc_{i+a\hat{y}}+c_i^\dag\sigma_zc_{i+a\hat{z}})+\text{H.c.}\\
&+\sum_i B(k_0^2-\frac{6}{a^2})c_i^\dag\sigma_zc_i+A_1(\frac{4}{a^2}-k_0^2)c_i^\dag c_i\\
&-\sum_i\frac{A_1}{a^2}(c_i^\dag c_{i+a\hat{x}}+c_i^\dag c_{i+a\hat{y}})+\text{H.c.}.
\end{split}
\end{equation}

The lattice constant is set to unity, $a=1$, and all physical parameters are chosen to be dimensionless. Two transport schemes with setups shown in Figs. 2(a) and 2(d) in the main text are studied.

(1) For the vertical spin transport in the $z$-direction, the lattice is built into the junction in Fig. 1(a) in the main text. The cross-section of the junction in the $x-y$ plane has the size $20a\times20a$. Spin polarized current is injected, and the final spin current is contributed by the incident and reflected electrons. The interface barrier $U=Z\sqrt{C|\bm{k}_F|Bk_0}$ is simulated by the on-site potential at the interface monolayer of the junction.

(2) For the lateral charge transport, the cross-section of the junction in the $x-y$ plane has the size $20a\times20a$, and the thickness of the normal metal in the $z$-direction is $15a$. Two ferromagnetic terminals are attached to the normal metal, with the same width and thickness. The chemical potentials in the ferromagnetic terminals are set to zero, $\mu_0=0$. Two terminals have opposite Zeeman terms $\pm M\sigma_z$, with a spin splitting of $M=4$.

For both minimal model and HgCr$_2$Se$_4$ model, the geometrical parameters of the setups are the same.

\subsection{Effect of interface disorder}
\begin{figure}
\centering
\includegraphics[width=0.6\textwidth]{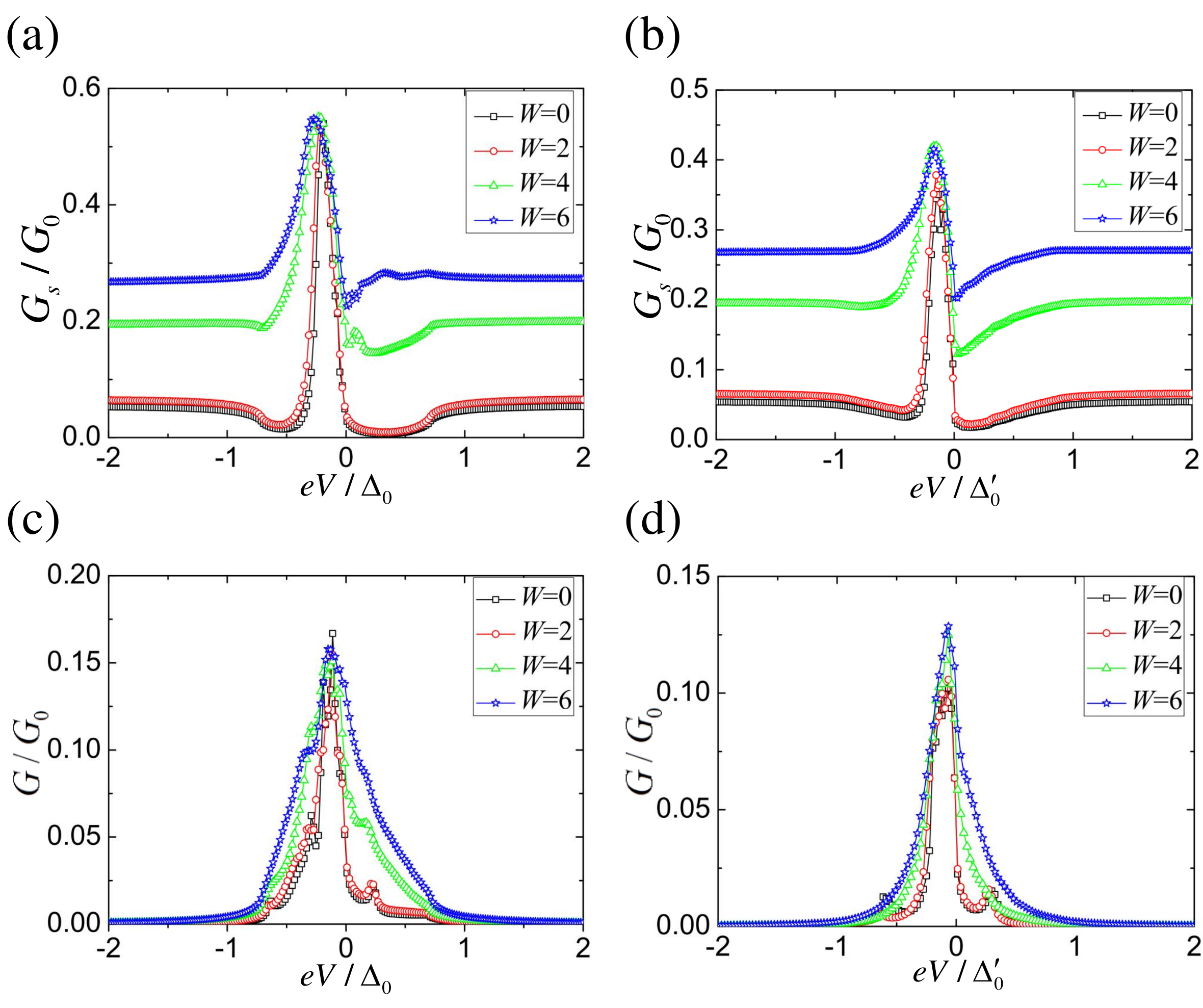}
\caption{(Color online). Interface disorder effect on the spin (a, b) and charge (c, d) transport schemes. (a) and (c) are calculated by the minimal model; (b) and (d) are calculated by the HgCr$_2$Se$_4$ model. The interface barrier is $Z=4$, and all the other parameters are the same as those in the main text.} \label{figs2}
\end{figure}

For real samples, interface disorder may exist. We simulate this effect by introducing uncorrelated Gaussian disorder with strength $W$ (in the same unit of $U$) to the interface monolayer. The results for both transport schemes, and both minimal and HgCr$_2$Se$_4$ models are shown in Fig.~\ref{figs2}. One can see that the resonant peaks are robust against the disorder.

\subsection{Effect of imperfect spin injection}
Without loss of generality, we assume the injected spin to be oriented in the $z$-direction. Imperfect spin injection can be treated as a weighted initial spin state $a=(a_\uparrow, a_\downarrow)^{\text{T}}=(\cos\frac{\theta'}{2}e^{i\varphi'},\sin\frac{\theta'}{2})^{\text{T}}$ with a distribution function $f(\theta',\varphi')$. The spin conductance is then calculated by a weighted average over $f(\theta',\varphi')$. We assume $f(\theta',\varphi')=f(\theta')/(2\pi)$ is independent of $\varphi'$ so that the net spin polarization is still in the $z$-direction. Below the energy gap $\Delta(\bm{k}_\parallel)$ where spin-flipped reflection dominates the spin transport, the incident spin state $a=(a_\uparrow, a_\downarrow)^{\text{T}}$ and reflected spin state $b=(b_\uparrow, b_\downarrow)^{\text{T}}$ are related by the unitary reflection matrix as $b^{\text{T}}=Ra^{\text{T}}$, with $R(\bm{k}_\parallel,E)=\left(
                                       \begin{array}{cc}
                                         r_c & r_f \\
                                         -r_f & r_c^*r_f/r_f^* \\
                                       \end{array}
                                     \right)$.
The spin conductance is contributed by all $\bm{k}_\parallel$ channels with an average over the injected spin state, yielding
\begin{equation}
\begin{split}
G_s&=\frac{e^2}{h}\frac{\mathcal{A}}{(2\pi)^2}\int\int dk_xdk_y \int_0^\pi d\theta'\int_0^{2\pi} d\varphi'f(\theta',\varphi')\text{Tr}\Big[a^*(\sigma_z-R^\dag\sigma_z R)a\Big]\\
&=\frac{e^2}{h}\frac{\mathcal{A}}{(2\pi)^2}\int\int dk_xdk_y \int_0^\pi d\theta' f(\theta')\cos\theta'(1+R_f-R_c)\\
&=\Lambda\frac{e^2}{h}\frac{\mathcal{A}}{(2\pi)^2}\int\int dk_xdk_y(1+R_f-R_c);\ \ \ \ \ \
\Lambda=\int_0^\pi d\theta' f(\theta')\cos\theta'.
\end{split}
\end{equation}
As a result, the effect of the imperfect spin injection is an overall prefactor $\Lambda$, which will not change the signature of RSFR. For a perfect spin injection, $f(\theta')=\delta(\theta')$, and the spin conductance reduces to Eq. (3) in the main text.

\end{document}